# The Grossberg Code: Universal Neural Network Signatures of Perceptual Experience

**Birgitta Dresp-Langley**

Centre National de la Recherche Scientifique UMR 7357 CNRS, Strasbourg University, FRANCE; birgitta.dresp@cnrs.fr

**\*** Correspondence: birgitta.dresp@cnrs.fr; Tel.: +33-388119117

**Abstract**

Two universal functional principles of Grossberg's Adaptive Resonance Theory [19] decipher the brain code of all biological learning and adaptive intelligence. Low-level representations of multisensory stimuli in their immediate environmental context are formed on the basis of *bottom-up activation* and under the control of *top-down matching* rules that integrate high-level long-term traces of contextual configuration. These universal coding principles lead to the establishment of lasting brain signatures of perceptual experience in all living species: from *aplysiae* to primates. They are re-visited in this concept paper here on the basis of examples drawn from the original code and from some of the most recent related empirical findings on contextual modulation in the brain, highlighting the potential of Grossberg's pioneering insights and groundbreaking theoretical work for intelligent solutions in the domain of developmental and cognitive robotics.

**Keywords:** multisensory perception; brain representation; contextual modulation; adaptive resonance; biological learning; self-organization; matching rules; winner-take-all principle;

1. Introduction

In his latest book [1], Grossberg discusses empirical findings and his own neural network models to illustrate, and forecast, how autonomous adaptive intelligence [2] is or may be implemented in artificial systems at unprecedentedly high levels of brain function [3,4,5]. His account of how the brain generates conscious cognition and, ultimately, individual minds provides mechanistic insights into complex phenomena such as mental disorders, or the biological basis of morality and religion. The author's theoretical work clarifies why evolutionary pressure towards adaptation and behavioral success not only explains the brain, but is also a source for model solutions to large-scale problems in machine learning, technology, and Artificial Intelligence. Adaptive brain mechanisms [6] are the key to autonomously intelligent algorithms and robots. They may be pre-determined by a universal developmental code, or "engram", that is channeled through the connectome by specific proteins/peptides embedded within pre-synaptic neuronal membranes [7] and corresponds to information provided by the electrical currents afferent to pre-synaptic neurons [8,9,10]. Grossberg's book [1] conveys a philosophical standpoint on shared laws of function in living systems, from the most primitive to the most advanced, showing how neurons support unsupervised adaptive learning in all known species, and how such biological learning has enabled the emergence of the human mind across the evolutionary process. Bearing this in mind, the present concept paper draws from the beginnings of this journey into the mind, which is described by Grossberg's significant early work on neural processes for perception, perceptual learning, and memory, aimed at understanding how the brain builds a cognitive code of physical reality. Since perception is the first step through which a brain derives sense from the raw data of a physical environment, his account for how elementary signals in the physical environment are processed by the neural networks of the brain was a mandatory achievement for understanding how inner representations of the outside world may be generated [11,12]. The ability to derive meaning from complex sensory input requires the integration of information over space and time as well as memory mechanisms to shape that integration [13] into contents of experience. In mammals with intact visual systems, this relies on processes in the primary visual cortex of the brain [14], where neurons integrate visual input along shape contours into neural association fields [15]. The geometric selectivity of ensembles of functionally dedicated neural networks is progressively fine-tuned by contextual modulation and experience towards long-term memory representation of all the different configurations likely to be encountered in natural scenes. Horizontal cortical connections provide a broad domain of potential associations in this process, and top-down control functions dynamically gate these associations to task switch the function of a given network [16]. Grossberg's work has provided a unified model of brain learning where horizontal cortical connections provide a broad range of potential, functionally specific neural associations through a mechanism called bottom-up activation [17], as will be explained and illustrated on the basis of examples. Mechanisms of adaptive resonance and top-down matching [17] then explain how the contextual modulation of visual and other sensory input drives dynamic brain learning to gate the links within and between neural association fields towards increasingly complex memory representations [16,18]. This concept paper uses two of the functional principles of Adaptive Resonance Theory [19] to illustrate the implications for unsupervised brain learning and adaptive intelligence. The examples chosen here is drawn from the original models and from related empirical findings. These are revisited under the light of some of the most recent advances in a conceptual discussion aimed at highlighting the potential of Grossberg's pioneering insights and groundbreaking theoretical work for intelligent solutions to some of the most difficult current problems in Artificial Intelligence (AI) and robotics. The following sections will elaborate on the biological principles of multisensory contextual modulation in the brain, in section 2, to illustrate the relevance of adaptive resonant learning as conceptualized in the Grossberg code, the functional principles of which are then explained further in section 3. Section 4 provides a generic ART system with its mathematical definition and an example of neural network architecture that could be implemented on this basis for autonomous and self-organizing multiple event coding to help control object-related aspects of environmental uncertainty in robotics.

**2. Contextual Modulation in the Brain**

The brain processes local information depending on the context in which this information is embedded. The representation of contextual information peripheral to a salient stimulus is critical to an individual's ability to correctly interpret and flexibly respond to stimuli in the environment. The processes and circuits underlying context-dependent modulation of stimulus-response function have mostly been studied in vertebrates [20], yet, well-characterized connectivity patterns are already found in the brains of lower level species such as insects [21], providing circuit-level insights into contextual processing. Recent studies in flies have revealed neuronal mechanisms that create flexible and highly context-dependent behavioral responses to sensory events relating to threats, food, and social interaction. Throughout brain evolution, functional building blocks of neural network architectures, with increasingly complex functional architecture have emerged across species, with increasingly complex long-range connectivity ensuring information encoding in processing streams that are anatomically segregated at a cellular level. The functional specificity of individual streams, long-range interactions beyond the classic receptive field of neurons and interneurons [22], and cortical feedback mechanisms [21, 23] provide an excellent model for understanding the complex processing characteristics inherent to individual streams as well as the extent and mechanisms of their interaction in the genesis of brain representation. Contextual modulation in the sensory cortex coding for vision, hearing, somatosensation and olfaction is partly under central control by the prefrontal cortex, as shown by some of the most recent evidence from neuroscience.

*2.1. Vision*

To be able to extract structure, form, and meaning from intrinsically ambiguous and noisy physical environments, the visual brain has evolved neural mechanisms dedicated to the integration of local information into global perceptual representations. This integration is subject to contextual modulation [23]. Mechanisms with differential sensitivity to relative stimulus orientation, size, relative position, contrast, polarity, and color operate within specific spatial scales to integrate local visual input into globally perceived structure [24,25,26,27]. The differential contextual sensitivity to color and luminance contrast in visual contextual modulation involves the luminance sensitive pathways (M-pathways) and the color sensitive pathways (P-pathways) of the visual brain [23,28] in a from-simple-to-complex-cells processing hierarchy at the level of the visual cortex, already predicted in Grossberg's early models of visual form representation [29,30]. The cooperative and competitive interactions between co-activating or mutually suppressive detectors in functionally dedicated neural networks suggested in the model were confirmed several years later in psychophysical and electrophysiological studies, taking into account response characteristics of orientation-selective visual cortical neurons as a function of the context in which visual target stimuli were presented [22,24,25]. Contextual modulation translates into effects where nearby visual stimuli either facilitate or suppress the detection of the targets (behavior), and increase or decrease the firing rates of the cortical neurons responding to the targets (brain). The cooperative and competitive brain-behavior loops depend on the geometry of so-called "perceptive fields" [22] within a limited range of size-distance ratios. The shorter temporal windows of achromatic context effects compared with chromatic contextual modulation [23,27] Cooperative mechanisms of contextual modulation in vision are subject to substantial practice (perceptual learning) effects, where top-down signals dynamically modulate neural network activities as a function of specific perceptual task constraints. Such top-down mediated changes in cortical states reflect a general mechanism of synaptic learning [4,8], potentiating or suppressing neural network function(s) depending on contextual relevance.

*2.2. Hearing*

Sounds in natural acoustic environments possess highly complex spectral and temporal structures, spanning over a whole range of frequencies, and with temporal modulations that differ within frequency bands. The auditory brain has the ability to reliably encode one and the same sound in a variety of different sound contexts, and to tell apart different sounds within a complex acoustic scene. Processing acoustic features such as sound frequency and duration is highly dependent on co-occurring, acoustic and other, sources of stimulation [32], and involves interactions between the external spectral and temporal context of an auditory target, and internal behavioral states of the individual such as arousal or expectation. Current findings suggest that sensory attenuation and neuronal modulation may happen during behavioral action as a consequence of disrupted memory expectations in the case of unpredictable concurrent sounds [33]. The auditory system demonstrates nonlinear sensitivity to temporal and spectral context, often employing network-level mechanisms, such as cross-band and temporally adaptive inhibition, to modulate stimulus responses across time and frequency [32]. How the auditory system modulates responses to sensory and behavioral contexts is not yet understood. The superior colliculus (SC) is a structure in the mammalian midbrain that contains visual and auditory neural circuits. In mice [34], auditory pathways from external nuclei of the inferior colliculus (IC) with direct inhibitory connections, and excitatory signals driving feed-forward inhibitory circuits within the SC were found. A previously unrecognized pathway, the lateral posterior nucleus (LP) of the thalamus projects extensively to sensory cortices. Bidirectional activity modulations in LP or its projection to the primary auditory cortex (A1) in awake mice reveal that LP improves auditory processing by sharpening neuronal receptive fields and their frequency tuning [35]. LP is strongly activated by specific sensory signals relayed from the superior colliculus (SC), contributing to the maintenance and enhancement of sound signal processing in the presence of auditory background noise and threatening visual stimuli respectively. This shows that multisensory bottom-up pathways plays a role in contextual [36] and cross-modality modulation of auditory cortical processing in mammals. Cross-modality modulation of sensory perception is necessary for survival. In a natural environment, organisms are constantly exposed to a continuous stream of sensory input depending on the environmental context. The response properties of neurons, dynamically adjust to contextual changes across all sensory modalities, and at different stages of processing from periphery to cortex.

*2.3. Somatosensation*

Cross-modality modulation implies that coincident non-auditory (visual, tactile) processing influences the neural networks underlying contextual modulation of hearing, or that non-visual (auditory, tactile) signals may reach the neural networks underlying the contextual modulation of vision. Touch has a direct effect on visual spatial contextual processing, for example [37]. Contextual modulation and neuronal adaptation in visual and auditory systems interact with sensory adaptation in the somatosensory system, but through which pathways and mechanisms is not yet well understood. The ability to integrate information from different sensory modalities is a fundamental feature of all sensory neurons across brain areas, which makes sense under the light of the fact that visual, auditory, and tactile signals originate from the same physical object when actively manipulated. The synthesis of multiple sensory cues in the brain improves the accuracy and speed of behavioral responses [38]. Task-relevant visual, auditory and tactile signals are experienced together in motor tasks [39], and pioneering work in neurophysiology from the 1960ies has shown convergence of visual, auditory, and somatosensory signals at the level of the pre-frontal cortex in cats [40]. Also, visual signals can bypass the primary visual cortex to directly reach the motor cortex, which is immediately adjacent and functionally connected to the somatosensory cortex [41]. Effects of neuronal adaptation on response dynamics and encoding efficiency of neurons at single cell and population levels in the whisker-mediated touch system in rodents illustrate that sensory adaptation provides context-dependent functional mechanisms for noise reduction in visual processing [42]. Between integration and coincidence detection, cross-modality modulation achieves energy conservation and disambiguates the encoding of principal features of tactile stimuli. Sensory systems do not develop and function independently. Early loss of vision, for example, alters the coding of sensory input in primary somatosensory cortex (S1) to promote enhanced tactile discrimination. Neural response modulation in S1 of mammals (opossums in this case) after elimination of visual input through bilateral enucleation early in development reveal neural origins of tactile experience in naturally occurring patterns of exploratory behavior after vision loss [43]. In early blind animals, overall levels of tactile experience were similar to those of sighted controls, and their locomotion activity was unimpaired and accompanied by normal whisking. Early blind animals exhibit a reduction in the magnitude of neural responses to whisker stimuli in S1, combined with a spatial sharpening of the neuronal receptive fields. The increased selectivity of S1 neurons in early blind animals is reflected by improved population coding of whisker stimulus positions, particularly along the axis of the snout aligned with the primary axis of the natural whisker motion. These findings suggest that a functionally distinct form of tactile (somatosensory) plasticity occurs when vision is lost early in development. After sensory loss, compensatory behavior mediated through the spared senses is generated through recruitment of brain areas associated with the deprived sense. Alternatively, functional compensation in spared modalities may be achieved through a combination of plasticity in brain areas corresponding to both spared and deprived sensory modalities.

*2.4. Olfaction*

Multisensory interactions in the brain need are most strongly relied upon and, therefore, need to be optimal when the stimulus ambiguity in a physical environment is highest [44]. Sensorial as well as central cross-modal signaling mechanisms contribute bottom-up and top-down contextual signaling. For example, both whisking and breathing are affected by the presence of odors in rodents, and the odors bi-directionally modulate activity in a small but significant population of the barrel cortex neurons through distinct bottom-up and top-down mechanisms [45]. In the human brain, different aspects of olfactory perception in space and time have been identified by means of EEG recordings [46] . Sensorial (low-level) representations of smell expand into larger areas associated with emotional, semantic, and memory processing in activities significantly associated with perception. These results suggest that initial odor information coded in the olfactory areas evolves towards perceptual realization through computations (long-range mechanism) in widely distributed cortical regions with different spatiotemporal dynamics [47]. Specific brain structures act appear to form hubs for integrating local multisensory cues into a spatial framework [48] enabling short-term as well as long-lasting memory traces of odors, touch sensations, sounds and visual objects in different dynamic contexts. Contextual modulation in the brain thus explains how olfactory and other sensory inputs translate into diverse and complex perceptions such as the pleasurable floral smell of flowers or the aversive smells of decaying matter. The prefrontal cortex (PFC) plays an important role in this process. Recent evidence suggests that the PFC has dedicated neural networks that receive input from olfactory regions, and that the activity of these networks is coordinated on the basis of   selective attention producing different brain alert   states [49].

*2.5. Prefrontal control*

In the mammalian brain, information processing in specific sensory regions interacts with global mechanisms of multisensory integration under the control of the PFC. Emerging experimental evidence suggests that the contribution of multisensory integration to sensory perception is far more complex than previously expected [42,43]. Associative areas such as the prefrontal cortex, which receive and integrate inputs from diverse sensory modalities, not only affect information processing in modal sensory pathways through down-stream signaling, but also influence contextual modulation and multisensory processing (Fig. 1).

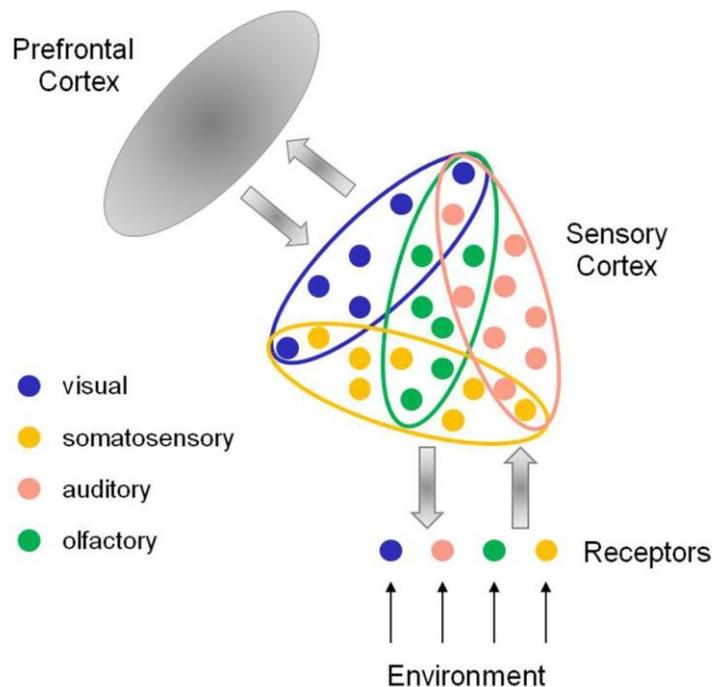

**Figure 1.** The prefrontal cortex receives and integrates signals from diverse sensory structures and pathways, and controls information processing in, and interaction between,

modal neural networks (visual, somatosensory, auditory, olfactory) through down-stream signaling.

Developmental mechanisms account for the interaction between the neuronal networks involved [50], with relevance for brain-inspired intelligent robotics, as will be discussed further later herein. In animals and humans, prefrontal downstream control is necessary in cases of conflicting sensory information, where signals from different modalities compete, or provide incongruent input data [51]. The brain then needs to reach a probabilistic decision on the basis of top-down control signals (perceptual experience). However, another remarkable ability of the brain is capacity to rapidly detect unexpected stimuli. Living beings depend on rapid detection of the unexpected when it is relevant (i.e. an alarm going off, for example) because it enables them to adapt behavior accordingly and swiftly. Prefrontal control also explains why irrelevant sounds are incidentally processed in association with the environmental context even though the contextual stimuli activate different sensory modalities [52]. This is consistent with brain data showing that top-down effects of the prefrontal cortex on contextual modulation of visual and auditory processing depend on selective attention to a particular sensory signal [53] among several coincident stimuli. Attempts to understand how functional interaction between different brain regions occurs through multisensory integration constitute a leading edge research area in contemporary neuroscience [54]. Low-level brain representation of information is not enough to explain how we perceive the world. To enable us to recognize and adaptively act upon objects in the physical world, lower-level sensory network representations need to interact with higher-level brain networks capable of coding contextual relevance.

**3. Brain signatures of perceptual experience**

How the brain generates short and long-term memory signatures of perceptual experience, and which mechanisms permit to retrieve and update these traces regularly during life-long brain learning and development (ontogenesis), is still not fully known. Well before contextual modulation and context-sensitive neural mechanisms were identified in neural circuits of different species, Grossberg had understood that they must exist and, considering the principles of unsupervised synaptic (Hebbian) learning [8], which had been demonstrated in low-level species such as *aplysia* [55], that they would have to be universal. In his early work on adaptive resonance [19], he proposed universal functional principles for the generation of short-term and long-term memory traces and their activation in context-sensitive processes of retrieval. These functional principles exploit two mechanisms of neural information processing in resonant circuits of the brain, referred to as bottom-up automatic activation and top-down matching.

*3.1. Bottom-Up Automatic Activation*

Bottom-up automatic activation is a mechanism for the processing and the temporary storage of perceptual input in short-term and working memory. Through bottom-up automatic activation, a group of cells within a given neural structure becomes potentiated, and is eventually activated, when it receives the necessary bottom-up signals. These bottom-up signals may or may not be consciously experienced. They are then multiplied by adaptive weights that represent long-term memory traces and influence the activation of cells at a higher processing level. Grossberg [17] originally proposed Bottom-Up Automatic Activation to account for the way in which pre-attentive processes generate learning in the absence of top-down attention or expectation. It appears that this mechanism is equally well suited to explain how subliminal signals may trigger supraliminal neural activities in the absence of phenomenal awareness [56,57]. Learning in the absence of phenomenal awareness accounts for visual statistical learning in newborn infants [58], and non-conscious visual recognition [59], for example. Bottom-up automatic activation may generate supraliminal brain signals, or representational contents with weak adaptive weights, as a candidate mechanism to explain how the brain manages to subliminally process perceptual input [60] that is either not directly relevant at a given moment in time, or cannot be made available to conscious processing because of a local brain lesion [59]. Grossberg [9,12,17,19] suggested that bottom-up activation may automatically activate target cell populations at higher levels of processing, as in bottom-up activation of the PCF by sensory cortices [47,49], for example.

*3.2. Top-Down Matching*

Top-down expectations are needed to consolidate traces of bottom-up representation through mechanisms that obey three properties: 1) they select consistent bottom-up signals and 2) suppress inconsistent bottom-up signals. Together these properties initiate a process that directs attention to a set of critical features that are consistent with a learned expectation. However, 3) a top-down expectation by itself cannot fully activate target cells. It can only sensitize, modulate, or prime the cells to respond more easily and vigorously if they are matched by consistent and sufficiently strong (relevant) bottom-up inputs. Were this not the case, we would hallucinate events that are not really there by mere top-down expectation. Top-down expectations therefore do not activate, only modulate representations, as discussed here above in *3.4*. Top-down representation matching is a mechanism for the selective matching bottom-up short-term or working memory representations to already stored and consolidated (learnt) memory representations (Fig. 2). Subliminal bottom-up representations may become supraliminal when bottom-up signals or representations are sufficiently relevant at a given moment in time to activate statistically significant top-down matching signals [60]. These would then temporally match the bottom-up representations (coincidence). A positive match confirms and amplifies ongoing bottom-up representation, whereas a negative match invalidates ongoing bottom-up representation. Top-down matching is a selective process where subliminal representations become embedded in long-term memory structures and temporarily accessible to recall, i.e. a conscious experience of remembering.

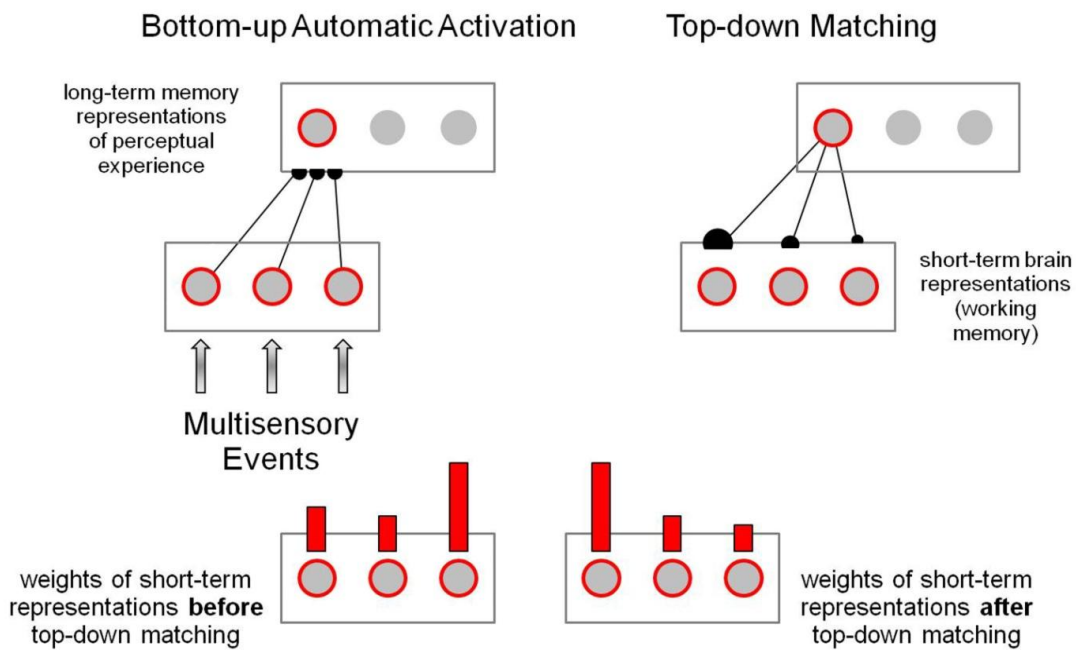

**Figure 2.** In the ART matching rules, bottom-up signals from the environment activate short-term memory representations in working memory which then, in turn, send bottom-up signals towards a subsequent processing stage at which long-term memory representations are temporarily activated (top left). These bottom-up signals are multiplied by learned long-term memory traces which selectively filter short-term representations and activate top-down expectation signals (top right) that are matched against the selected representations in working

memory. The strength of the matches determines the weighting of short-term representations (bottom) after top-down matching.

*3.3. Temporary representation for selection and control*

Grossberg's universal coding rules produce temporary and long-term brain signatures of perceptual experience. They address what he called the attention-pre-attention interface problem [9,12,17,19] by allowing pre-attentive (bottom-up) processes to use some of the same circuitry that is used by attentive (top-down) processes to stabilize cortical development and learning. Consistently, research on human cognition [61] has confirmed that attention ensures the selection of contents in working memory, controlled by mechanisms of filtering out irrelevant stimuli and removing no-longer relevant representations, while working memory contributes to controlling perceptual attention as well as action by holding templates available for perceptual selection and action sets available to implement current goals [61]. Top-down matching in its most general sense generates feed-back resonances between bottom-up and top-down signals to rapidly integrate brain representations and hold them available for a consciousness experience at a given moment in time. Non-conscious semantic priming is explained on these grounds. Statistically significant positive top-down matching signals produced on the basis of strong signal coincidences explain why subliminal visual representations become conscious when presented in a specific context, especially after a certain amount of visual learning or practice [60]. Conversely, significant negative matches produced on the basis of repeated discrepancies generating strong negative coincidence signals could explain why a current conscious representation is suppressed and replaced by a new one when a neutral conscious representation is progressively and consistently weakened by association with a strongly biased representation, as in evaluative conditioning and contingency learning [57,58]. Some of the above mentioned functional properties require long-range connectivity in cortical circuits capable of generating what Edelman [62] called "reentrant signaling". Bottom-up representations that activate specific structures of such circuits, but do not produce sufficiently strong matches to long-term memory signals, will remain non-conscious [60]. Strong positive top-down matching of selected representations will compete with weaker or negative matches and, ultimately, be suppressed from conscious experience like, for example, in cases where the conscious integration of new input interferes with the conscious processing of anything else [35,50]. Specific instructions telling subjects what to look for, or what to attend to, in a visual scene may generate top-down expectation signals strong enough to inhibit matching of other relevant signals at the same moment in time [31]. Top-down matching generates neural computations of event coincidence [63]. Results from certain observations in motor behavior without awareness [64] highlight potential implications of negative top-down matching for conscious control in learning. Individuals may become aware of unconsciously pursued goals of a motor performance or action when the latter does not progress well, or fails. This could reflect the consequence of repeated negative top-down matching of the non-conscious bottom-up goal representation and top-down expectation signals in terms of either memory traces of previous success, or representations of desired outcome. Repeated and sufficiently strong negative matching signals might thereby trigger instant consciousness of important discrepancies between expectancy and reality [65]. Awake mammals can switch between alert and non-alert brain states hundreds of times every day. The effects of alertness on two cell classes in layer 4 of primary visual cortex, excitatory "simple" cells and fast-spike inhibitory neurons, show that for both cell classes, alertness increases their functional (excitatory or suppressive) strength, and considerably enhances the reliability of visual responses [66]. In simple cells, alertness increases the temporal frequency bandwidth, but preserves contrast sensitivity, orientation tuning, and selectivity for direction and spatial frequency. Alertness selectively suppresses the simple cell responses to high-contrast stimuli and stimulus motion orthogonal to their preferred direction of movement. This kind of conscious feed-back control fulfills an important adaptive function, and has evolved in response to the pressures of intrinsically ambiguous and steadily changing physical environments. The mathematical development and equations describing

ART resonant learning it its most generic form were made explicit in the Cohen–Grossberg model [67,68], which will be detailed further here below with respect to the development of adaptive intelligence in robotics.

**4. Towards adaptive intelligence in robotics**

Resonant brain states are a key concept of ART. They arise from the self-organizing principles of biological neural learning whereby our brains autonomously adapt to a changing world. Biological neural learning, unlike the learning algorithms that fuel Artificial Intelligence, is driven by evolution, with a remarkable pressure towards increasingly higher levels consciousness across the phylogenesis [69]. Pressure towards the development of increasingly autonomous and adaptively intelligent forms of agency also exists in the growing field of robotics, in particular neurorobotics [70]. Detailed descriptions and equations describing the full span of potential for the development of autonomously intelligent robots may be found in [71,72,73,74]. The most generic functional principles of ART are aimed at was been termed the hierarchical resolution of uncertainty. Hierarchical resolution of uncertainty means that multiple processing stages are needed for brains to generate sufficiently complete, context-sensitive, and stable perceptual representations upon for successful action by intelligent agents. The mathematical development and equations describing ART resonant learning it its most generic form are inspired by the principles of Hebbian neural (synaptic) learning, and are given by the Cohen–Grossberg model [67,68]. The latter is defined in terms of the following system of nonlinear differential equations describing interactions in time $t$ among and between neural activities $x_i$, or short-term memory (STM) traces, of any finite number of individual neurons or neuronal populations (networks)

$$dx_i/dt = a_i(x_i)\,[b_i(x_i) - \sum_j c_{ij} d_j(x_j)] \qquad (1)$$

with symmetric interaction coefficients $c_{ij} = c_{ji}$ for weak assumptions of state-dependent non-negative amplification functions $a_i(x_i)$, self-signaling functions $b_i(x_i)$, and competitive interaction functions $d_j(x_j)$. Magnitudes for $i, j = 1, 2, . ., n$ and $n$ may be chosen arbitrarily. Each population in (1) can have its own functions $a_i(x_i)$, $b_i(x_i)$, and $d_j(x_j)$. One possible physical interpretation of the symmetric interaction coefficients $c_{ij} = c_{ji}$ is that the competitive interactions depend upon Euclidean distances between the populations. Defined as in (1), the $i$th population activity x can only grow to become momentarily a "winner" of the competition at times $t$ where the competitive balance $[b_i(x_i) - j\,c_{ij} d_j(x_j)] > 0$. When $[b_i(x_i) - j\,c_{ij} d_j(x_j)] < 0$, the given population is "losing" the competition. The ART-inspired neural network architecture for multiple event coding, represented schematically here above, can be implemented by exploiting properties and parameters of the system described in (1). This would permit implementing robot intelligence with capacities beyond reactive behavior. The selective filtering of relevant sensory input from a multitude of external inputs, and to autonomously generate adaptive sequences of memory steps to identify and recognize specific visual objects in the environment, permits to control external perturbations acting on a robot–object system. This is possible in a system like the one illustrated here above on the sole basis of internal dynamics of the resonant network. The ability to correctly identify objects despite multiple changes across time is a competence required in many engineering applications that interact with the real world such as robot navigation. Combining information from different sensory sources promotes robustness and accuracy of place recognition. However, mismatch in data registration, dimensionality, and timing between modalities remain challenging problems in multisensory place recognition [75]. We may, as ART stipulates, define intelligence as the ability to efficiently interact with the environment and to plan for adequate behavior based on the correct interpretation of sensory signals and internal states.

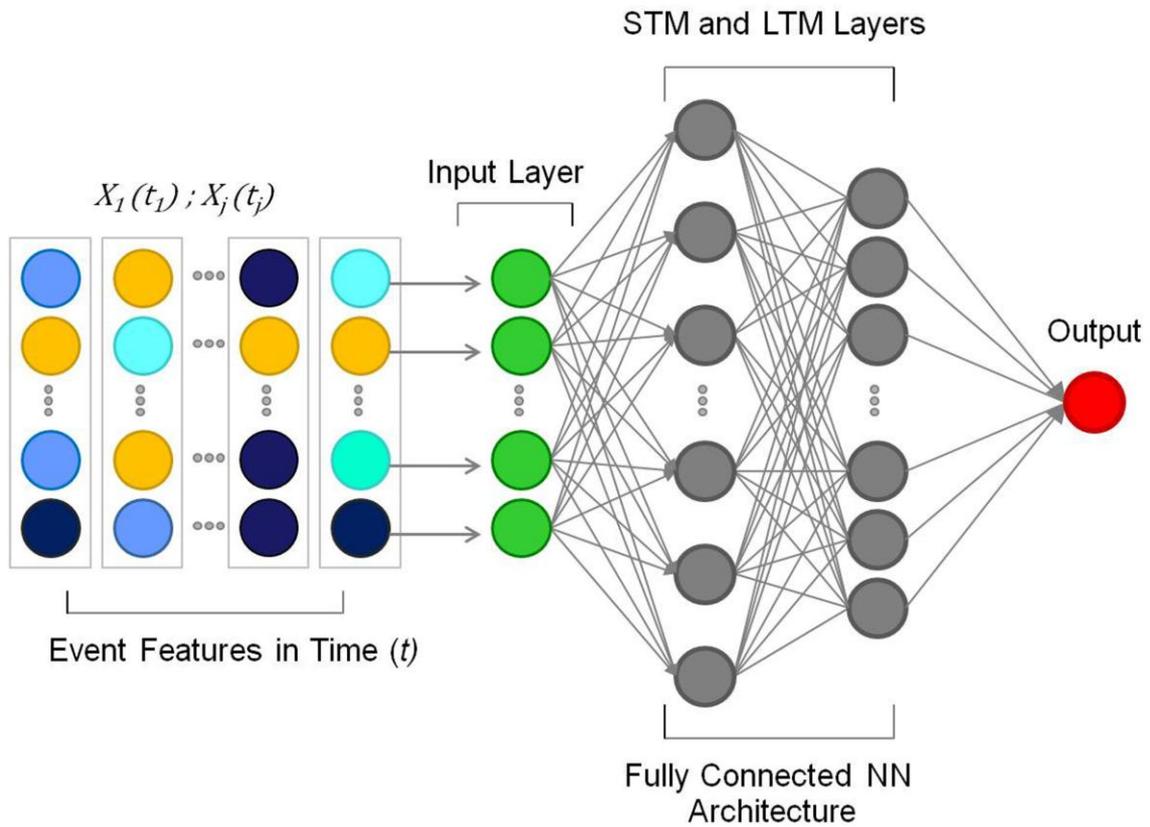

**Figure 3.** ART-inspired Neural Network architecture for adaptively intelligent event coding across time *t*.

This means that an intelligent agent or robot will be successful in accomplishing its goals, able to learn and predict the effects of its actions, and to continuously adapt to changes in real-world scenarios. Ultimately, embodied intelligence allows a robot to interact swiftly with the environment in a wide range of conditions and tasks [76]. The ART model made explicit here above in (1), a Hebbian-learning based and mathematically parsimonious system of non-linear equations, can be directly implemented to enable intelligent multi-event coding across time *t* (Fig.3) for robot control by adaptive artificial intelligence (neurons or neural populations).

## 5. Discussion

Grossberg's universal coding rules enable learning in non-stationary unexpected world, while classic machine learning approaches assume a predictable and controlled world [2]. Unlike passive adaptive filters [77], they enable self-organized unsupervised learning akin to biological synaptic learning [2,4,5,8,55]. The ART matching rules actively focus attention to selectively generate short and long-term brain signatures of critical features in the environment, which is achieved by dynamic, non-passive, steadily updated synaptic weight changes in the neural networks [9,12,17,19]. The top-down control of selective processing involves activation of all memory traces to match or mismatch bottom-up representations globally using *winner-takes-all* best- match criteria. Neural network architectures driven by the ART matching rules do not need labeled data to learn, as previously explained in [2]. In short, the Grossberg code overcomes many of the computational problems of back propagation and Deep Learning models. Equipping cognitive robots with artificial intelligence that processes and integrate cross-modal information according to such self-organized contextual learning ensures that they will interact with the environment more efficiently, in particular under conditions of sensory uncertainty [4,78]. The universal ART matching rules are directly relevant to a particular field in robotics that is motivated by human cognitive and behavioral development, i.e. developmental robotics. The goal is to probe developmental or environmental aspects of cognitive processes by exploring robotic capabilities for interaction using artificial sensory systems, and autonomous motor capabilities in challenging environmental platforms [79]. As illustrated here in this paper, low-level sensory and high-level neural networks interact in a bottom-up and top-down manner to create coherent perceptual representations of multisensory environments. Similarly, bottom-up and top-down interactions for the integration of multiple sensory input streams play a crucial role in the development of autonomous cognitive robots by endowing agents with improved robustness, flexibility, and performance. In cases of ambiguous or incongruent cross-sensory inputs, for example, biological inspiration acquires a major role. Autonomous robots with odor-guided navigation [80] can benefit from multisensory processing capabilities similar to that found in animals, allowing them to reliably discriminate between chemical sources by integrating associated auditory and visual information. Cross-modal interaction with top-down matching can enable the autonomous learning of desired motion sequences [81] matching expected outcomes from audio or video sequences, for example. Approaches to multisensory fusion in robotic systems directly inspired by the distributed functional architecture of the mammalian cortex have existed for some time [82]. Biological inspiration exploiting top-down cross-modal processing is mandatory for autonomous cognitive robots that acquire perceptual representations on the basis of active object exploration and groping. By actively processing geometric objet information during motor learning, aided by tactile and visual sensors, it becomes possible to reconstruct the shape, relative position, and orientation of objects. Service robotics is a fast-developing sector that requires embedded intelligence into robotic platforms that interact with humans and the surrounding environment. One of the main challenges in this field is robust and versatile manipulation in everyday life. Embedding anthropomorphic synergies into the gripper mechanical design [83] helps, but autonomous grasping still represents a challenge, which can be resolved by endowing robots with self-organizing multisensory adaptive capabilities, as discussed here above. Combining biological neural network learning with compliant end-effectors would not only permit optimizing the grasping of known deformable objects [84], but also help intelligent robots anticipate and grasp unforeseen objects. Bottom-up activation combined with top-down control gives robots the capability to progressively learn in an ever-changing multisensory environment by means of self-organizing interaction with the environment (Fig. 4). Implementing multisensory memories in robotics in such a way permits equipping intelligent agents with sensory-cognitive adaptive functions that enable the agents to cope with the unexpected in complex and dynamic environments [85]. Lack of multisensory perceptive capabilities, on the other hand, compromises continuous learning of robotic systems because internal models of the multisensory world can then not be acquired and adapted throughout development.

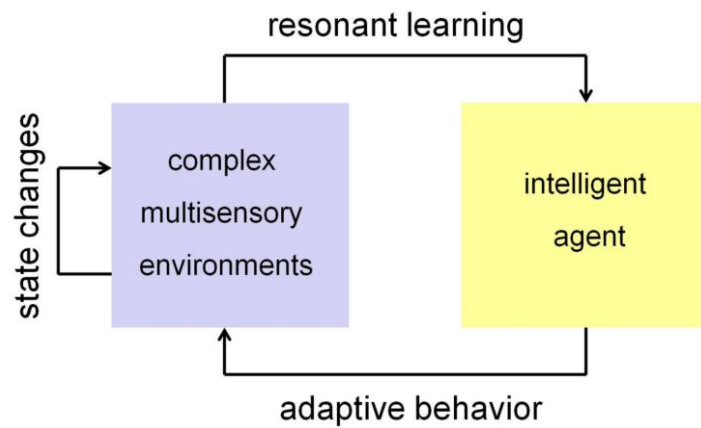

**Figure 4.** Adaptive resonance as a universal principle of biological learning gives intelligent agents the capability to cope and evolve with ever-changing multisensory environments on the basis of self-organizing adaptive behavior (interaction with the environment).

Adaptive resonance is a powerful concept that provides model approaches for a multitude of human interactions. The relationship between the physical mechanism of resonance and its biological significance in the genesis of perceptual experience in neural networks across all species, from mollusks to humans, makes it also a powerful concept for human–robot interaction, at all functional levels and within a wider cultural and scientific context. Resonant brain states, established on the basis of matching processes involving top-down expectation and bottom-up activation signals, drive all biological learning at lower and higher levels. Learning in biological neural networks is by nature unsupervised and best accounted for in terms of competitive *winner-takes-all* matching principles [86,87,88]. A resonant state is predicted to persist long enough and at a high enough activity level to activate long-term signatures of perceptual experience n dedicated neural networks. This explains how these signatures can regulate the brain's fast information processing, observed at the millisecond level, without any awareness of the signals that are being processed. Through resonance as a mediating event, the combination of universal matching rules and their attention focusing properties, learning and responding to arbitrary input environments becomes stable. In the mammalian brain, such stability may be reflected by the ubiquitous occurrence of reciprocal bottom-up and top-down cortico-cortical and cortico-thalamic interactions [89].

**6. Conclusions**

Well before contextual modulation and context-sensitive mechanisms were identified in neural circuits of different species, Grossberg had understood that they have to exist. The principles of unsupervised synaptic (Hebbian) learning had been demonstrated in low-level species such as *aplysia*, pointing towards universal principles of perceptual coding. In his earliest work on adaptive resonance, Grossberg set the foundations of universal functional principles of neural network learning for the generation of brain traces of perceptual experience, and their activation by context-sensitive, dynamic, self-organizing mechanisms producing resonant brain states. Equipping cognitive robots with artificial intelligence based on adaptive resonance, processing and integrating cross-modal information in self-organized contextual learning, will produce intelligent robots that interact with complex environments adaptively and efficiently, in particular under conditions of sensory uncertainty.

**Supplementary Materials:** All data and conceptual work discussed here are available in the material cited

**Funding:** This research received no external funding.

**Acknowledgments:** Material support from the CNRS is gratefully acknowledged

**Conflicts of Interest:** The author declares no conflict of interest.